\documentclass[twocolumn,prl,superscriptaddress,preprintnumbers,amsmath,amssymb]{revtex4-2}

\usepackage{graphicx}
\usepackage{float}
\usepackage{xspace}
\usepackage{xcolor}
\usepackage{amsmath}

\newcommand{\phii}{Institute of Physics II, University of Cologne, D-50937 Cologne, Germany}
\newcommand{\geo}{Section Crystallography, Institute of Geology and Mineralogy, University of Cologne, D-50674 Cologne, Germany}
\newcommand{\krecl}{K$_2$ReCl$_6$\@\xspace}
\newcommand{\kptcl}{K$_2$PtCl$_6$\@\xspace}
\newcommand{\kircl}{K$_2$IrCl$_6$\@\xspace}
\newcommand{\kirbr}{K$_2$IrBr$_6$\@\xspace}
\newcommand{\ksncl}{K$_2$SnCl$_6$\@\xspace}
\newcommand{\ag}{A$_\text{1g}$\@\xspace}
\newcommand{\eg}{E$_\text{g}$\@\xspace}
\newcommand{\tg}{T$_\text{2g}$\@\xspace}
\newcommand{\cf}{\emph{cf}.\@\xspace}

\newcommand{\fig}[1]{Fig.\@\xspace\ref{#1}}

\graphicspath{{../figures/}}

\begin{document}

\title{Local symmetry breaking and low-energy continuum in K$_2$ReCl$_6 $}

\author{P.~Stein}
  \affiliation{\phii}
\author{T.~C.~Koethe}
  \affiliation{\phii}
\author{L.~Bohat\'y}
  \affiliation{\geo}
\author{P.~Becker}
  \affiliation{\geo}
\author{M.~Gr\"uninger}
	\affiliation{\phii}
\author{P.~H.~M.~van~Loosdrecht}
  \affiliation{\phii}



\begin{abstract}
\noindent Using polarization selective spontaneous Raman scattering, we have investigated the 5$d$ transition metal compound K$_2$ReCl$_6$ which displays a series of structural phase transitions. We observe a violation of the Raman selection rules in the cubic high temperature phase as well as a low-energy scattering continuum persistent throughout the investigated temperature range from 300 down to 5~K. The continuum couples to one of the phonon modes at temperatures above the lowest structural phase transition at 76~K. 
We propose a common origin of these observations caused by disorder in the orientation of the ReCl$_6$ octahedra which locally breaks the long-range cubic symmetry. Consistent results from the related non-magnetic compound \ksncl support this interpretation. 
\end{abstract}


\maketitle

\section{Introduction}

The material class of composition $A_2M$(Cl,Br)$_6$ where 
$A$ is typically an alkaline metal and $M$ a tetravalent main group element or transition metal, 
features a signi\-fi\-cant variety of structural and magnetic properties \cite{ReviewArticle}. The parent compound \kptcl has a stable cubic antifluorite structure ($\mathrm{Fm\bar{3}m}$), with $\mathrm{PtCl_6} $ octahedra on a face centered cubic (fcc) lattice and potassium ions occupying the tetrahedral voids, see Fig.\,\ref{Sample_Structure_Phases}. Many members of the family display a series of crystallographic phase transitions as a function of temperature \cite{BuseyBr,NH3_transition,Os_transition,Se_transition,K2SnCl6}. A prime example of this class is \krecl, which in its high temperature phase has the cubic structure of its parent compound.  
\krecl shows three structural phase transitions at $T_\text{c3}=111$ K, $T_\text{c2}=103$ K and $T_\text{c1}=76$~K, all of which are easily observable in thermodynamic measurements \cite{BuseyCV,Willemsen}, along with a long-range ordered antiferromagnetic phase below $T_\text{N}=12$~K \cite{BuseyX,SmithBacon}. Early reports have claimed that all structural phase transitions are driven by soft modes corresponding to consecutive rotations of the
octahedra around different axes \cite{OLeary,JWLynn,111K_angle}. Upon cooling below $T_\text{c3}$, the structure becomes tetragonal with minimal deviations from the cubic phase. The low-temperature phases below $T_\text{c2}$ possess monoclinic symmetry. Remarkably, the correct crystallographic structure of the tetragonal phase and the nature of the octahedral rotations have been determined only very recently \cite{Alexandre}.
Lately, the strong interest in unconventional phenomena driven by spin-orbit interactions has spawned renewed attention for this material class, in part motivated by recent theoretical studies which have proposed a spin-orbit assisted Jahn-Teller activity for certain materials of this family \cite{Khomskii}. In particular, the $t_{2g}^3 $ configuration of \krecl (Re: $5d^3$) was suggested to be a potential realization.   

Previous Raman measurements on this compound at ambient and liquid helium temperatures have identified the Raman active phonon modes \cite{OLeary}. The appearance of additional modes at the phase transitions reflects the lowering of the symmetry, and soft rotary modes have been identified \cite{OLeary,JWLynn} as the driving force behind these transitions. In this work, we investigate \krecl using temperature dependent Raman spectroscopy in two orthogonal polarization configurations extending to lower energies than previously reported. We observe that the cubic symmetry is already broken locally in the high temperature phase. 
Furthermore, we reveal a low-energy excitation continuum and discuss its relationship to the local symmetry breaking observed at high temperature. 

\section{Experimental}

\begin{figure}[t]
	\centering
	\includegraphics[width=.45\textwidth]{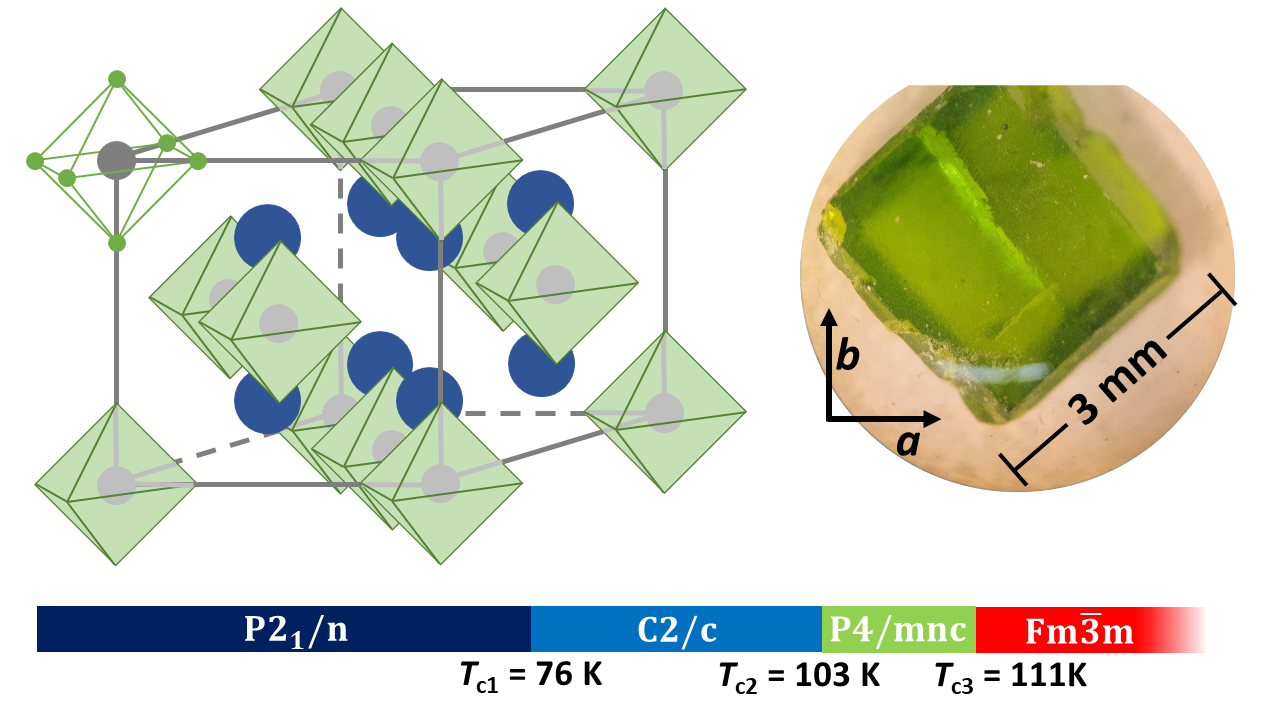}
	\caption{Left: Schematic illustration of the room temperature structure with space group $\mathrm{Fm\bar{3}m} $. Blue, grey and green spheres represent $\mathrm{K^+} $, $\mathrm{Re^{4+}} $ and $\mathrm{Cl^{-}} $ ions, respectively. Right: The (001) surface used for these measurements. The orientation of the cubic $a$ and $b$ axes is indicated. Bottom: The different crystallographic phases and transition temperatures of \krecl within the temperature range covered.}
	\label{Sample_Structure_Phases}
\end{figure}

High-quality transparent single crystal samples of \krecl were grown from HCl solution by controlled slow evaporation of the solvent and subsequently characterized by x-ray diffraction, specific heat and susceptibility measurements \cite{Alexandre}. A picture of the sample can be seen in Fig.\,\ref{Sample_Structure_Phases}.
The sample was mounted in a liquid helium flow cryostat. Measurements were performed on a polished (001) surface in backscattering geometry with the polarization of incoming light parallel to the cubic \emph{a} axis using a 532 nm solid state laser with a power of  0.5~mW at the sample and a spot size of 30 $\mathrm{\mu m}$. The total integration time was eight minutes per temperature step in the range from 5 K up to room temperature.

Spectra were recorded using a triple grating spectrometer equipped with a liquid nitrogen cooled CCD camera yielding a total energy resolution well below 1 meV. The phonon energies were determined by applying a multi-peak fit using Lorentzian and Fano line shapes, where appropriate.

Crystal growth and Raman measurements of \ksncl were done analogously with a slightly lower laser power due to stronger scattering as compared to \krecl. 

\section{Results}

\begin{figure}[t]
	\centering
	\includegraphics[width=.45\textwidth]{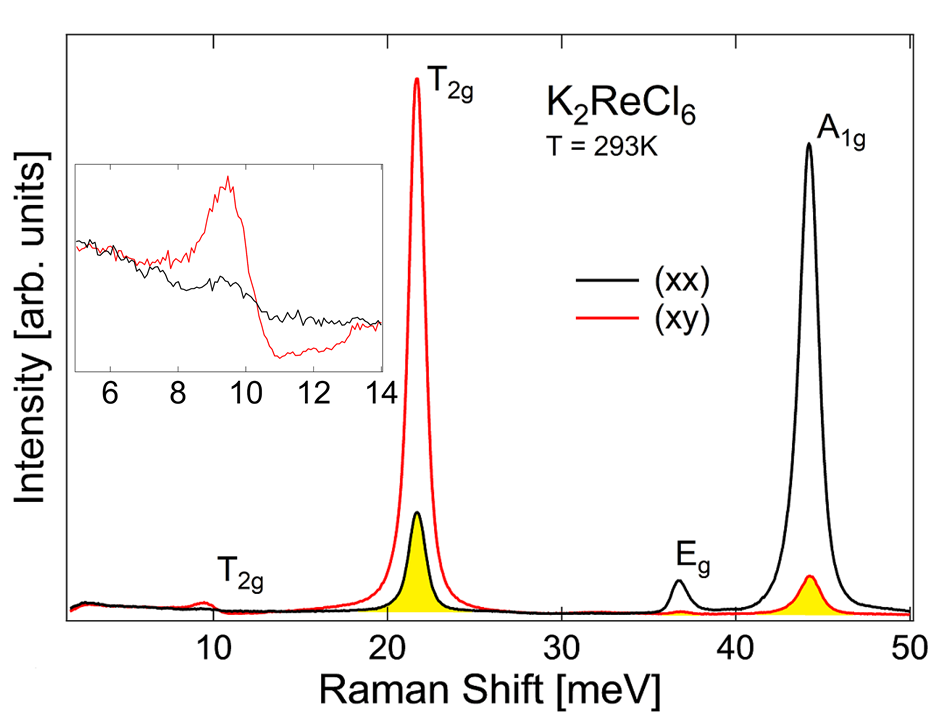}
	\caption{Raman spectra of $\mathrm{K_2ReCl_6} $ at $T$ = 293 K in $(xx)$ (black) and $(xy)$ (red) polarization configurations. Yellow areas indicate the violation of cubic selection rules. The inset shows a blow-up of the low-energy region containing the lattice phonon.}
	\label{cubic_Spectrum}
\end{figure}
The vibrational spectrum of \krecl is split into two regions. The internal vibrations of the octahedra sit in the higher energy part of the spectrum above 15~meV, while the external lattice modes have energies below 12 meV. The irreducible representation of the Raman active modes in the high-temperature cubic phase is $\Gamma^{\prime}  = 1\text{A}_\text{1g}  + 1\text{E}_\text{g}  + 2\text{T}_\text{2g} $, where one of the T$_\text{2g}$ modes is the external potassium mode and the remaining three modes are internal vibrations of the octahedra.
By symmetry, the selection rules predict that in our experimental geometry, the A$_\text{1g} $ and E$_\text{g}$ modes are detectable exclusively in the $(xx)$ polarization configuration, where only the component of scattered light with the polarization vector parallel to that of the incoming light is analyzed, while the two T$_\text{2g}$ modes only appear in the $(xy)$ polarization configuration, where the polarization vectors of incoming and analyzed light are perpendicular. 
The Raman spectrum of the room temperature phase in these two configurations is shown in Fig.\,\ref{cubic_Spectrum}. 
The four Raman active modes can be seen at the energies of 10 meV, 22 meV, 37 meV and 44~meV, respectively, in agreement with the values reported in previous Raman scattering experiments \cite{OLeary} and density of states calculations \cite{DOS}.

The comparably weak external phonon mode at $\approx $10 meV is shown 10 times expanded in the inset of Fig.\,\ref{cubic_Spectrum}. It has a strongly asymmetric Fano line shape, which indicates an interaction of the vibrational excitation with a scattering continuum \cite{Fano}. Indeed, such a continuum is present in our data at energies reaching up to at least 12 meV. We note that previous reports have failed to resolve this low-energy continuum due to less efficient suppression of elastic scattering from the excitation light source.

As marked by the yellow areas in Fig.\,\ref{cubic_Spectrum}, the cubic selection rules are not perfectly fulfilled in this crystal. The spectrum taken in $(xy)$ configuration contains both the \ag and \eg modes with about 5\% of their intensity in the $(xx)$ configuration. Similarly, the \tg modes appear in the $(xx)$ spectrum with 10\% of their intensity from the $(xy)$ polarization. Since the sample has been oriented using Laue diffraction, and both the experimental alignment and polarization of the light have been carefully verified, so-called leakage of a scattering signal into the wrong polarization channel due to misalignment or insufficient polarization can be ruled out. Moreover, the Raman tensor of the \ag mode is rotationally invariant, therefore it is excluded to observe this mode in $(xy)$ configuration for cubic symmetry. 


\begin{figure}[t]
	\centering
	\includegraphics[width=\linewidth]{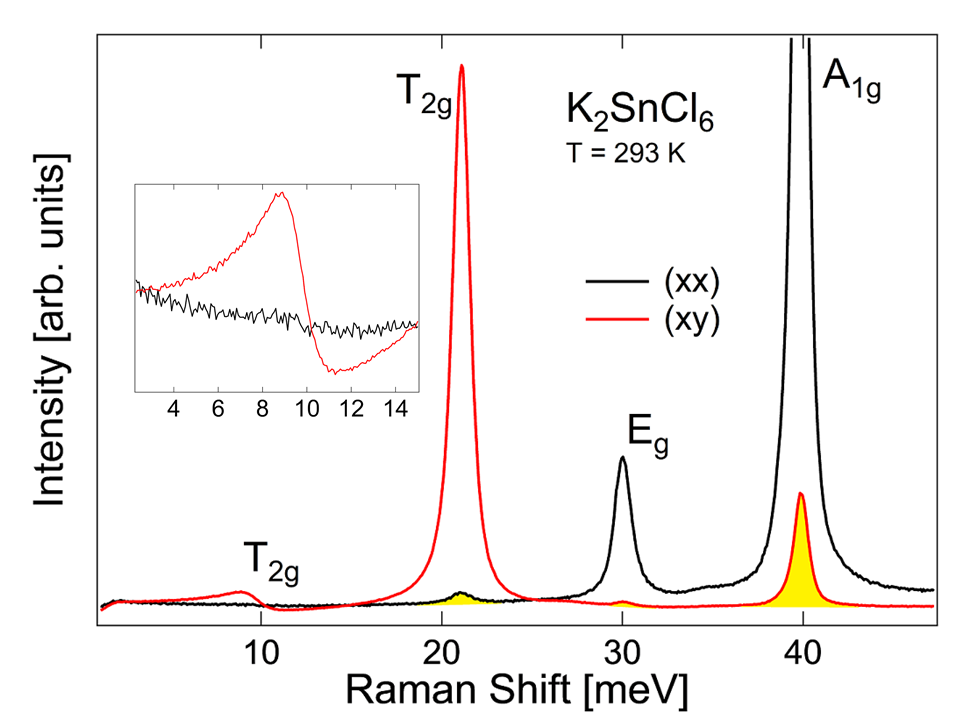}
	\caption{Room temperature Raman spectra of \ksncl in $(xx)$ (black) and $(xy)$ (red) polarization configuration, \cf \fig{cubic_Spectrum}.
	}
	\label{fig:k2sncl6}
\end{figure}

\begin{figure*}[t]
	\centering
	\includegraphics[width=0.45\textwidth]{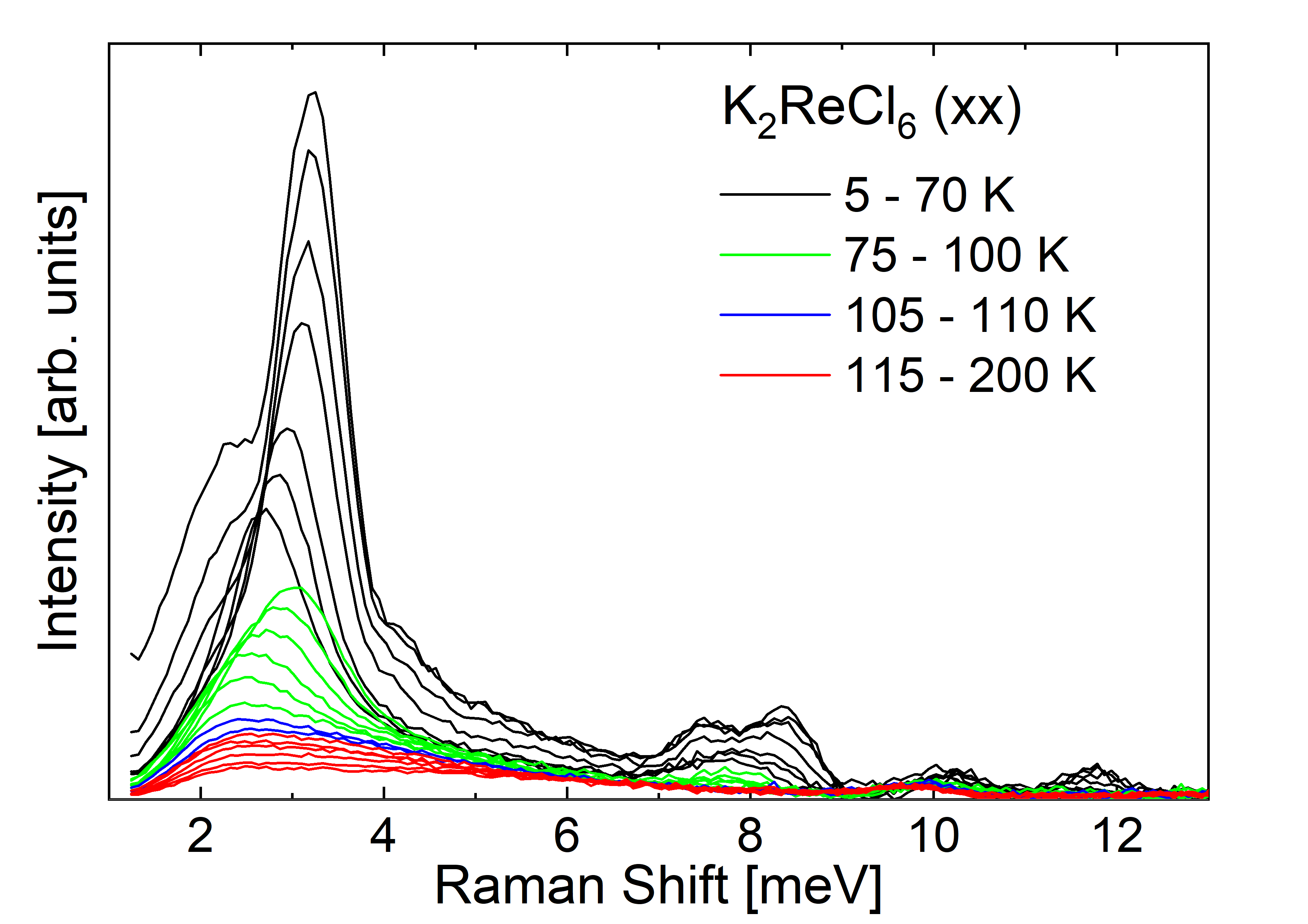}\hspace{4mm}
	\includegraphics[width=0.45\textwidth]{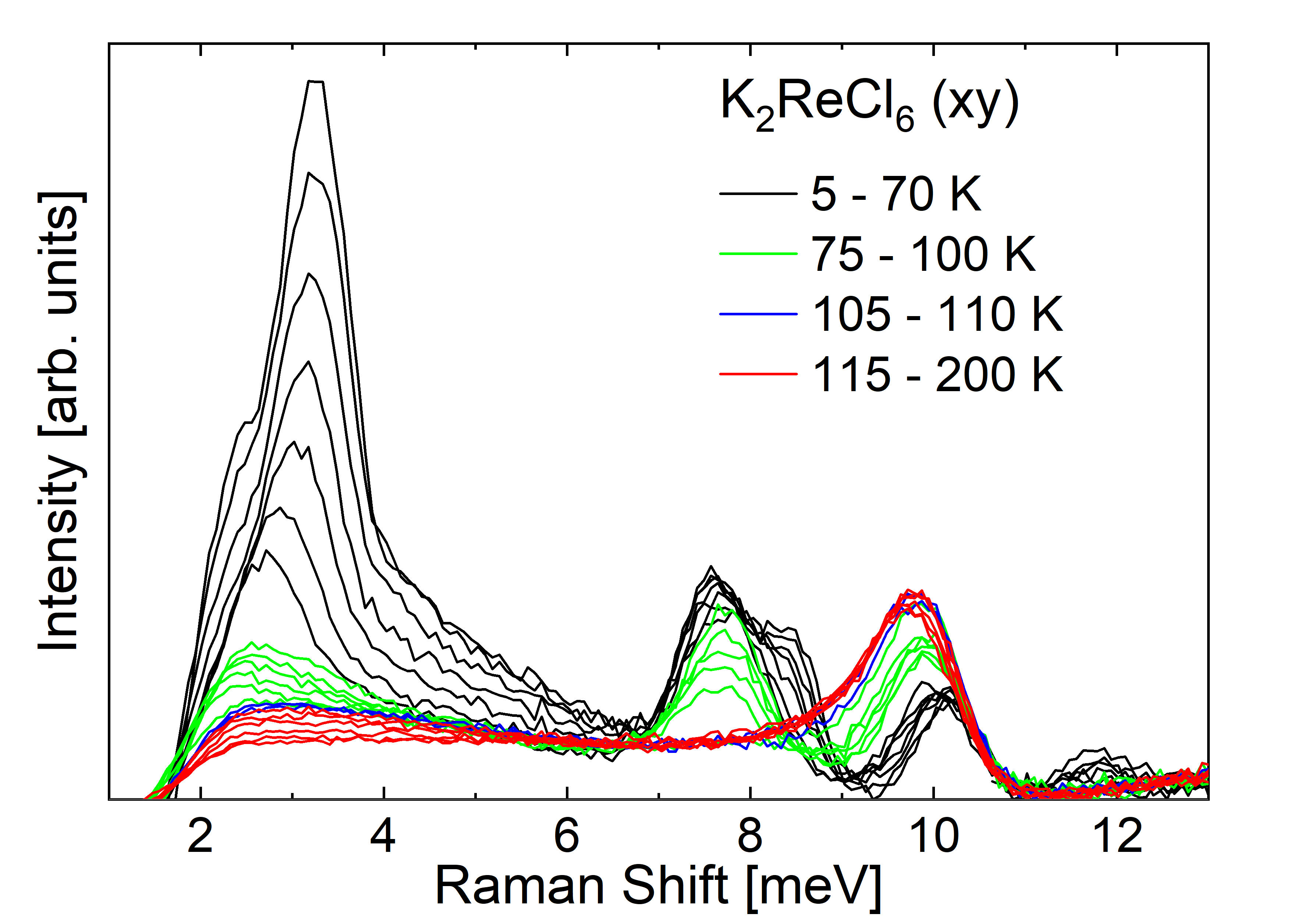}
	\caption{Bose-factor corrected Raman spectra of \krecl in the energy region below 13 meV at temperatures between 5 and 200 K; Left: ($xx$) polarization configuration; Right: ($xy$) polarization configuration. The different colors represent the four distinct crystallographic phases.}
	\label{Tdep}
\end{figure*}

In order to test whether or not the experimental observations might be related to the electronic or magnetic properties of the material, we have investigated the related non-magnetic compound \ksncl, where Sn has a filled shell $4d^{10}$ configuration. The room temperature Raman spectra for both the $(xx)$ and $(xy)$ configurations shown in \fig{fig:k2sncl6} reveal a violation of the cubic selection rules as well as a low-energy scattering continuum, consistent with the results of \krecl. We thus conclude that the origin of these observations must be related to the properties of the lattice structure rather than its electronic or magnetic nature. The same features can also be observed in a spectrum of \krecl measured with an excitation wavelength of 488 nm (not shown), which excludes that these features are caused by resonance scattering effects.


We now turn to the temperature dependence of the Raman spectra of \krecl. 
We focus on the low energy region below 13 meV shown in \fig{Tdep} for both ($xx$) (left) and ($xy$) (right) polarisation configurations. To extract the bare scattering strength, a constant offset has been subtracted from the raw data, and the prefactor $(n + 1)$ of the scattering cross section, with the energy and temperature dependent Bose occupation $n$, has been corrected for. Despite the lowering of the symmetry at the transition from the cubic (shown in red) to the tetragonal (blue) phase, no additional phonons are detected below $T_\text{c3}=111$~K, in agreement with previous reports on several materials in this family \cite{Raman1,Raman2,Raman3,Raman4,Raman5,Raman6}. We infer that the small degree of deformations in the tetragonal phase results in a minute energy splitting of modes which is insignificant in comparison to the lifetime and therefore not observed. 

\begin{figure}[t]
	\centering
	\includegraphics[width=0.45\textwidth]{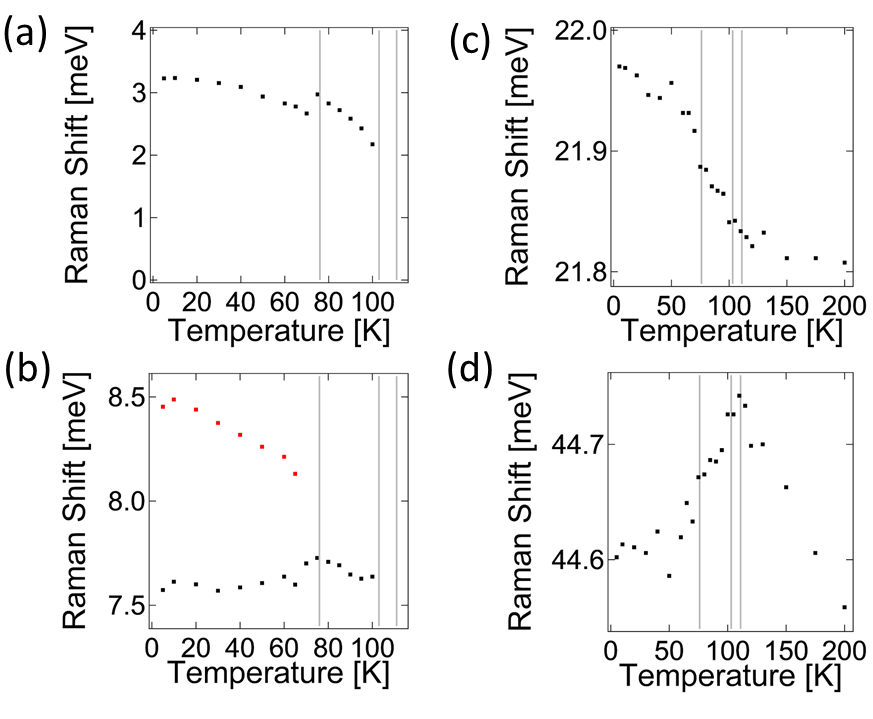}
	\caption{Temperature dependence of phonon energies: (a) The low-energy rotary mode; (b) The monoclinic modes near 8 meV; (c) The octahedral \tg mode; (d) The octahedral \ag mode. Vertical bars indicate phase transition temperatures. 
	}
	\label{Tdep_E_octa}
\end{figure}

The transition from tetragonal to the monoclinic C2/c structure at $T_\text{c2}=103$~K is clearly marked by two additional modes in the spectra (green). At 7.6 meV a reasonably strong mode emerges below $T_\text{c2} $ in both polarisation configurations, though more pronounced in ($xy$). Around 3 meV, another mode grows out of the low energy continuum in ($xx$) configuration. Right below the transition, the 3~meV mode is rather weak and thus difficult to distinguish from the strong scattering continuum, but it gains weight and definition towards lower temperatures. 
Below $T_\text{c1}=76$~K, the mode also appears in ($xy$) configuration, \cf black spectra in \fig{Tdep} (right).
As can be seen in Fig.\,\ref{Tdep_E_octa}\,(a), its excitation energy shows a weak anomaly at $T_\text{c1}$, and it softens substantially as the temperature approaches $T_\text{c2}$ from below. It is thus identified as the rotary mode previously proposed to drive the phase transition \cite{Alexandre,OLeary,JWLynn}.

In the low-temperature monoclinic $\mathrm{P2_1/n} $ phase below $T_\text{c1}=76$~K (shown in black in Fig. \ref{Tdep}), a distinct mode appears at 12 meV, and yet another one splits off of the phonon at 7.6 meV. 
The latter one again gains weight towards lower temperatures and shows significant softening towards $T_\text{c1}$, see Fig.\,\ref{Tdep_E_octa}\,(b). 
It is worth mentioning that the low-temperature spectrum agrees with previous reports \cite{OLeary}, with the notable exception of the aforementioned scattering continuum observed at low energies. 
%
%


The temperature dependence of the energies of the octahedral \tg and \ag phonons, shown in Figs.\,\ref{Tdep_E_octa}\,(c) and (d), respectively, reveals anomalies at the phase transitions. However, a splitting of the octahedral 
modes is not observed (not shown), indicating the absence of a sizable deformation of the octahedra down to the lowest temperatures. 
 
We now take a closer look at the cubic lattice mode at 10 meV which involves vibrations of the potassium ions, \cf \fig{Tdep_Fano}. The asymmetric line shape hints at a coupling of the mode with a continuous excitation band, and can be described by the Fano model \cite{Fano}, 

\begin{equation*}
I  = I_0  \frac{(q + \varepsilon)^2}{1 + \varepsilon^2}
\end{equation*}

with the Fano parameter $q$ and the reduced energy $\varepsilon $. The coupling strength between the phonon and the continuum, which can be quantified by 1/$q^2 $, is substantial and essentially constant above $T_\text{c2}$, but drops rapidly below 100 K and vanishes below $T_\text{c1} = 76$ K, see Fig.\,\ref{Tdep_Fano}\,(b). This is directly visible in the Raman spectra as the mode gets more and more symmetric when the temperature is lowered (Fig.\,\ref{Tdep_Fano}\,(a)).
The temperature dependence of the model parameter $\varepsilon$ corresponding to the bare phonon energy is depicted in Fig.\,\ref{Tdep_Fano}\,(c). It  shows anomalies at the same temperatures: the mode hardens down to approximately $T_\text{c2}$, then its energy decreases, before it begins to rise again below $T_\text{c1}$. 

%
\section{Discussion}

Our main experimental results are threefold:\\ 1.~The Raman selection rules in the high-temperature cubic phase are violated;\\ 2.~Below 12 meV, we observe a wide scattering continuum that shows an unusual temperature dependence and persists up to 300 K;\\
3.~The vibrational mode of the cubic phase involving motions of the $\mathrm{K^+} $ ions strongly couples to this continuum. As we will discuss in the following, all three observations are consistently explained by a local breaking of the cubic symmetry due to randomly rotating $\mathrm{ReCl_6} $ octahedra.

The cubic crystal structure in the high-temperature phase of this material, as well as many other members of the \kptcl family, has been well established experimentally using x-ray and neutron diffraction techniques \cite{Aminoff,JWLynn,Alexandre}. 
%
Anomalously high values of the atomic displacement parameters and finite
librational angles 
of the octahedra 
have been reported previously for several members of this material class, including \krecl \cite{Grundy,K2IrCl6_tilt,K2IrBr6_tilt}. Corresponding disorder was recently found for the isostructural compounds \kirbr and \kircl \cite{K2IrBr6_tilt,K2IrCl6_tilt,LeeK2IrCl6}. This leads us to the conclusion that the local symmetry of the octahedra deviates from the average symmetry which is retained on large length scales. The origin of this disorder may be dynamic fluctuations, static displacements, chlorine isotope disorder (natural abundance 75.8\% $^{35}$Cl, 24.2\% $^{37}$Cl), or a combination of these.
Being sensitive to the local symmetry, Raman scattering will reveal such local deviations from the global symmetry through apparently modified selection rules. 

\begin{figure}[t]
	\centering
	\includegraphics[width=.44\textwidth]{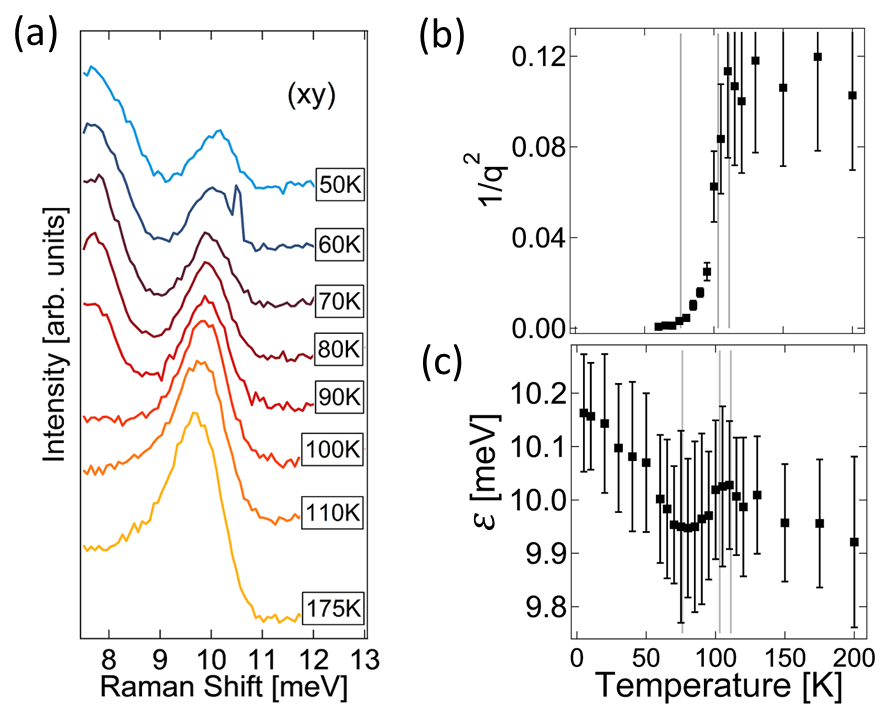}
	\caption{Temperature dependence of the external lattice mode: (a) Raman spectra from 50 to 175 K; (b) The coupling strength between phonon and continuum represented by $1/q^2$; (c) Reduced energy $\varepsilon$. Vertical bars indicate phase transition temperatures.
	}
	\label{Tdep_Fano}
\end{figure}

Here, it is instructive to recall a peculiarity of the crystal structure of the \kptcl family. 
The lattice can be imagined to be derived from the perovskite structure $AB$O$_3$ with half of the $B$ sites not occupied. The stability and stiffness of the perovskite structure originates to a large part from the interlocking of corner-sharing octahedra which requires any local deformation to propagate throughout the lattice, as is commonly observed in many perovskite materials \cite{Glazer72}.
However, with half of the $B$ site ions missing in the lattice of the \kptcl family, the octahedra disconnect and the structure loses much of its rigidity. In fact, the filled octahedra are even substantially compressed \cite{Grundy} and can thus be considered as only weakly interacting objects placed on the face-centered cubic positions of the antifluorite lattice. Therefore, the energy potential for rotations of the octahedra must be unusually flat and strongly anharmonic. 
 
%
Such an anharmonic potential shape gives rise to low-lying excitations connected to fluctuations in both the tilt angles and axes, resulting in a low-energy scattering continuum, as observed in Raman scattering.

As shown in \fig{Tdep}, the scattering strength of the continuum decreases continuously with increasing temperature, which is unexpected. The presence of several phonon modes overlapping with the continuum prevents a quantitative determination of the continuum spectral weight. 
Note that such local fluctuations of the octahedral orientation would not show up in the previously mentioned phonon density of states calculation \cite{DOS}. We conclude that the continuum spectral weight must be determined by the density of states of the low energy excitations which itself depends on the details of the potential landscape in which they exist. This requires a detailed analysis of the potential and its temperature dependence, which is beyond the scope of this study.

A similar low-energy continuum caused by fluctuations in the octahedral orientations has recently been observed in Raman scattering of perovskite-derived metal-organic hybrid materials \cite{Tilting,Sharma}. Interestingly, the observed fluctuations have been related to structural phase transitions in the material. 
It is important to note that the continuum in the case of \krecl is at best moderately 
affected by the structural phase transitions. 
The preferred average orientation of the octahedra changes across the structural phase transitions, but the fluctuations persist in a very similar way around the new average orientation.
This is consistent with previous claims that the structural phase transitions in this material class are purely displacive in nature and do not feature ordering of any kind, as an order-disorder transition would require a much higher entropy than that obtained in thermodynamic measurements \cite{ReviewArticle,BuseyCV}. Together with the fact that the continuum persists throughout the temperature range of our measurements, these observations indicate that the arrangement of the octahedra remains in a disordered state down to at least 5 K. This agrees with the idea of a very weak interaction between the octahedra, resulting in liquid-like properties. Future investigations of this phenomenon at the lowest experimentally accessible temperatures are highly desired.

In a recent publication \cite{LeeK2IrCl6}, a similar low-energy scattering feature that also shows increasing intensity towards lower temperatures was reported for the isostructural \kircl and identified as a central peak. As central peaks are a by-product of second-order structural phase transitions \cite{CentralPeak1,CentralPeak2,CentralPeak3,CentralPeak4}, this was seen as an indication for the existence of such a phase transition at very low temperatures. In our case, the continuous increase in continuum weight towards lower temperatures even across the two second-order structural phase transitions at $T_\text{c2} $ and $T_\text{c3} $ indicates that the continuum is unrelated to the possible spectral signatures of a central peak.

The asymmetric line shape of the cubic potassium mode at 10~meV above 76~K results from the significant coupling to the continuum. This  indicates a strongly anharmonic decay of the phonon, as has previously been reported for perovskite metal-organic hybrid materials \cite{Tilting,Sharma}.
%
Remarkably, this coupling nearly vanishes in the low-temperature monoclinic phase, while the scattering continuum persists, \cf Figs.\,\ref{Tdep}, \ref{Tdep_Fano}. 
As has been discovered very recently, the phase transitions at $T_\text{c2}$ and $T_\text{c3}$ are characterized by in-phase and out-of-phase rotations of the octahedra, respectively, while the phase transition at $T_\text{c1}$ is accompanied by a discontinuous change in the tilt axis of the octahedra \cite{Alexandre}. We infer that this is responsible for the observed suppression of the coupling between the potassium phonon mode and the continuum.  

Finally, we address another experimental observation which is considered to be unrelated to the main observations and their interpretation, yet worth mentioning. In contrast to previous reports that the octahedral phonon modes do not reflect the structural changes related to the crystallographic phase transitions \cite{OLeary,Raman1,Raman3,Raman6}, our spectra do reveal such signatures, although comparatively moderate, see Figs.\,\ref{Tdep_E_octa} (c) and (d). 
The failure to detect these has previously 
been taken as a proof that the ReCl$_6$ octahedra remain undistorted through all of the phase transitions. We believe that this conclusion still holds, despite the weak sensitivity of the octahedral modes. In particular, we observe no indication for substantial spin-orbit related Jahn-Teller distortions of the octahedra \cite{Khomskii}, which are expected to lead to a splitting of the degenerate octahedral phonons. 

Our interpretation is further supported by the observation that our three main results are also present in Raman spectra of \ksncl, see \fig{fig:k2sncl6}. In this material, Sn has a 4$d^{10} $ electronic configuration where both spin-orbital effects and Jahn-Teller interaction are absent. 

\section{Conclusion}

Polarization selective Raman spectroscopy was performed on a (001) surface of $\mathrm{K_2ReCl_6} $ in the temperature range between 5 K and room temperature. We observe a violation of the selections rules in the cubic high-temperature phase, together with a low-energy scattering continuum which shows unusual temperature dependence. 
We propose that both observations originate from disordered local tilts and rotations of the ReCl$_6$ octahedra. Thus the cubic symmetry is broken locally but remains intact on average over large length scales. Although this global structure changes at the phase transitions, the disorder survives down to temperatures of at least 5 K due to the very weak interaction between neighboring ReCl$_6 $ octahedra. The sharp signatures of the octahedral phonon modes in the Raman spectrum, the significant size of the thermal displacement ellipsoids \cite{PrivCom} and the increase of continuum weight for lower temperatures point to dynamic disorder as the origin of these phenomena, although a possible role of static disorder, e.g. caused by the different isotopes of Cl, cannot be fully excluded with our experimental methods.
The coupling between the potassium phonon mode and the low-energy scattering continuum results in anharmonic decay and gets suppressed at the transition to the low-temperature phase at $T_\text{c1}$. 
The persistence of the octahedral tilting down to low temperatures could be of great relevance to the magnetic interactions in the material, especially of the kind proposed for the sister compounds $\mathrm{K_2IrCl_6} $ and $\mathrm{K_2IrBr_6} $ \cite{K2IrCl6_tilt,K2IrBr6_tilt}. 

Lastly, we did not detect any spectral features indicative of a sizeable Jahn-Teller distortion that was suggested for this compound based on the large spin-orbit coupling in 5$d$ transition materials.\\

\begin{acknowledgments}
The authors would like to thank A.\ Bertin, M.\ Braden, D.\ Khomskii, T.\ Lorenz, F.\ Parmigiani, and P.\ Warzanowski for inspiring discussions. 
Funded by the Deutsche Forschungsgemeinschaft (DFG, German Research Foundation) -- Project number 277146847 -- CRC 1238. 
\end{acknowledgments}

\end{document}